\documentclass[envcountchap]{svmult}

\usepackage{mathptmx}       
\usepackage{helvet}         
\usepackage{courier}        
\usepackage{type1cm}        
\usepackage{makeidx}         
\usepackage{graphicx}        
\usepackage{multicol}        
\usepackage[bottom]{footmisc}

\makeindex             

\usepackage{amssymb,amsfonts,amsmath,bm,theorem,cite}

\newcommand{\bc}{\begin{center}}
\newcommand{\ec}{\end{center}}
\newcommand{\be}{\begin{equation}}
\newcommand{\ee}{\end{equation}}
\newcommand{\ba}{\begin{eqnarray*}}
\newcommand{\ea}{\end{eqnarray*}}
\newcommand{\bna}{\begin{eqnarray}}
\newcommand{\ena}{\end{eqnarray}}

\newcommand{\mpaa}{\begin{minipage}[t]{7.5cm}}
\newcommand{\mpea}{\end{minipage}}

\begin{document}

\title{Search for Food of Birds, Fish and Insects}
\author{Rainer Klages}
\institute{Rainer Klages \at Max Planck Institute for the Physics of
  Complex Systems, N\"othnitzer Str.~38, D-01187 Dresden, Germany and
  Queen Mary University of London, School of Mathematical Sciences,
  Mile End Road, London E1 4NS, UK, \email{r.klages@qmul.ac.uk}}
%
%
\maketitle

\abstract*{This book chapter introduces to the problem to which extent
  search strategies of foraging biological organisms can be identified
  by statistical data analysis and mathematical modeling. A famous
  paradigm in this field is the {\em L\'evy Flight Hypothesis}: It
  states that under certain mathematical conditions L\'evy flights,
  which are a key concept in the theory of anomalous stochastic
  processes, provide an optimal search strategy. This hypothesis may
  be understood biologically as the claim that L\'evy flights
  represent an {\em evolutionary adaptive} optimal search strategy for
  foraging organisms. Another interpretation, however, is that L\'evy
  flights {\em emerge} from the interaction between a forager and a
  given (scale-free) distribution of food sources. These hypotheses
  are discussed controversially in the current literature. We give
  examples and counterexamples of experimental data and their analyses
  supporting and challenging them.}

\section{Introduction}
  
\begin{figure}
\centerline{\includegraphics[width=8cm]{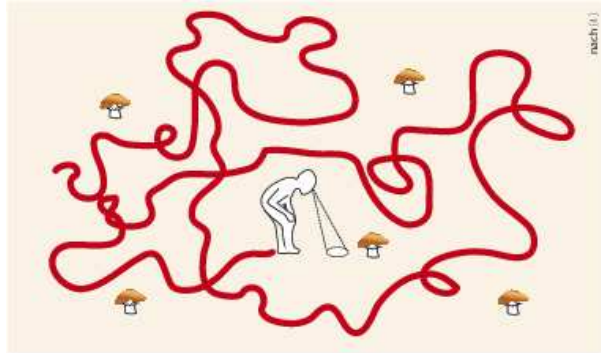}}
\caption{Illustration of a typical search problem \cite{CBV15,Kla15}:
  A human searcher endeavours to find mushrooms that are randomly
  distributed in a certain area. It would help to have an {\em optimal
    search strategy} that enables one to find as many mushrooms as
    possible by minimizing the search time.}
\label{fig:pilze} 
\end{figure}

\index{search}\index{search!strategy}
When you are out in a forest searching for mushrooms you wish to fill
your basket with these delicacies as quickly as possible. But how do
you {\em search efficiently} for them if you have no clue where they
grow (Fig.~\ref{fig:pilze})? The answer to this question is not only
relevant for finding mushrooms \cite{CBV15,Kla15}. It also helps to
understand how white blood cells kill efficiently intruding pathogens
\cite{HaBa12}, how monkeys search for food in a tropical forest
\cite{RFM03}, and how to optimize the hunt for submarines
\cite{Shles06}.

\index{operations research}\index{movement ecology} \index{L\'evy}
\index{L\'evy!Flight}\index{L\'evy!Flight!Hypothesis}
\index{L\'evy!stable distribution}
\index{L\'evy!motion}\index{foraging} In society the
problem to develop efficient search strategies belongs to the realm of
{\em operations research}, the mathematical optimization of
organizational problems in order to aid human decision-making
\cite{Stone07}. Examples are the search for landmines, castaways or
victims of avalanches. Over the past two decades {\em search research}
\cite{Shles06} attracted particular attention within the fields of
ecology and biology. The new discipline of {\em movement ecology}
\cite{Nath08,MCB14} studies foraging strategies of biological
organisms: Prominent examples are wandering albatrosses searching for
food \cite{Vis96,Vis99,Edw07}, marine predators diving for prey
\cite{Sims08,Sims10}, and bees collecting nectar \cite{LICCK12,LCK13}.
Within this context the {\em L\'evy Flight Hypothesis} (LFH) became
especially popular: It predicts that under certain mathematical
conditions on the type of food sources long {\em L\'evy flights}
\cite{SZK93} minimize the search time \cite{Vis96,Vis99,VLRS11}. This
implies that for a bumblebee searching for rare flowers the flight
lengths should be distributed according to a power law. \index{power law} 
Remarkably, the prediction by the LFH is completely different
from the paradigm put forward by Karl Pearson more than a century ago
\cite{Pea06}, who proposed to model the movements of biological
organisms by simple random walks as introduced in Chap.~2 of this
book. His suggestion entails that the movement lengths are distributed
exponentially according to a Gaussian distribution, see Eq.(2.10) in
this section.  L\'evy and Gaussian processes represent fundamental but
different classes of diffusive spreading. Both are justified by a
rigorous mathematical underpinning.

\index{central limit theorem}\index{central limit theorem!generalized}
More than 60 years ago Gnedenko and Kolmogorov proved mathematically
that specific types of power laws, called {\em L\'evy
  stable distributions} \cite{KRS08,ZDK15}, obey a central limit
theorem. Their result generalizes the conventional central limit
theorem for Gaussian distributions, which explains why Brownian motion
is observed in a huge variety of physical phenomena. But exponential
tails decay faster than power laws, which implies that for
L\'evy-distributed flight lengths there is a larger probability to
yield long flights than for flight lengths obeying Gaussian
statistics. Consequently, L\'evy flights should be better suited to
detect sparsely, randomly distributed targets than Brownian motion,
which in turn should outperform L\'evy motion when the targets are
dense. This is the basic idea underlying the LFH. Empirical tests of
it, however, are hotly debated
\cite{Edw07,Buch08,dJWH11,Pyke15,Reyn15}: Not only are there
problems with a sound statistical analysis of experimental data sets
when checking for power laws; their biological interpretation is also
often unclear: For example, for monkeys living in a tropical forest
who feed on specific types of fruit it is not clear whether the
observed L\'evy flights of the monkeys are due to the distribution of
the trees on which their preferred fruit grows, or whether the
monkeys' L\'evy motion represents an evolutionary adapted optimal
search strategy helping them to survive \cite{RFM03}.  Theoretically
the LFH was motivated by random walk models with L\'evy-distributed
step lengths that were solved in computer simulations \cite{Vis99}. A
rigorous mathematical proof of the LFH remains elusive.

This chapter introduces to the following fundamental question
cross-linking the fields of ecology, biology, physics and mathematics:
{\em Can search for food by biological organisms be understood by
  mathematical modeling?}  \cite{BLMV11,VLRS11,MCB14,ZDK15} It
consists of three main parts: Section~\ref{sec:lfh} reviews the LFH.
Section~\ref{sec:lonl} outlines the controversial discussion about its
verification by including basics about the theory of L\'evy motion.
Section~\ref{sec:bee} illustrates the need to go beyond the LFH by
elaborating on bumblebee flights. We summarize our discussion in
Sec.~\ref{sec:summ}.

\section{L\'evy motion and the L\'evy Flight Hypothesis}\label{sec:lfh}
  
\subsection{L\'evy flights of wandering albatrosses}

\begin{figure}
\centerline{\includegraphics[width=6cm]{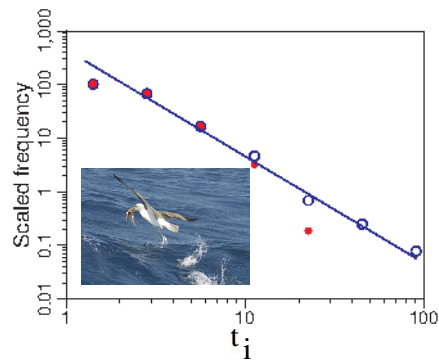}}
\caption{Histogram where `scaled frequencies' holds for the number of
  flight time intervals of length $t_i$ (in hours) normalized by their
  respective bin widths. The data is for five albatrosses during 19
  foraging bouts (double-logarithmic scale). Blue open circles show
  the data from Ref.~\cite{Vis96}. The straight line indicates a power
  law $\sim t^{-\mu}$ with exponent $\mu=2$. The red filled circles
  are adjusted flight durations using the same data set by eliminating
  times that the birds spent on an island \cite{Edw07}. The histogram
  is reprinted by permission from {\em Macmillan Publishers Ltd:
    Nature Ref.~\cite{Edw07}, copyright 2007}. The inset shows an
  albatross catching food; reprinted by permission from {\em Macmillan
    Publishers Ltd: Nature Ref.~\cite{Shles06}, copyright 2006}.}
\label{fig:alba} 
\end{figure}

\index{albatross}
In 1996 Gandhimohan Viswanathan and collaborators published a pioneering
article in the journal {\em Nature} \cite{Vis96}. For albatrosses
foraging in the South Atlantic the flight times were recorded by
putting sensors at their feet. The sensors got wet when the birds were
diving for food, see the inset of Fig.~\ref{fig:alba}. The duration of
a flight was thus defined by the period of time when a sensor remained
dry, terminated by a dive for catching food. The main part of
Fig.~\ref{fig:alba} shows a histogram of the flight time intervals of
some albatrosses. The straight line represents a L\'evy stable
distribution proportional to $\sim t^{-\mu}$ with an exponent of
$\mu=2$. By assuming that the albatrosses move with an on average
constant speed one can associate these flight times with a respective
power law distribution of flight lengths. This suggests that the
albatrosses were searching for food by performing L\'evy flights. 

For more than a decade albatrosses were considered to be the most
prominent example of an animal performing L\'evy flights.  This work
triggered a large number of related studies suggesting that many other
animals like deer, bumblebees, spider monkeys and fishes also perform
L\'evy motion \cite{Vis99,RFM03,Sims08,Sims10,VLRS11}.

\subsection{The L\'evy Flight Hypothesis}\label{sec:tlfh}

\index{L\'evy!Flight!Hypothesis}\index{search!strategy}
In 1999 the group around Gandhimohan Viswanathan published another
important article in {\em Nature} \cite{Vis99}. Here the approach was
more theoretical by posing, and addressing, the following general
question:

``What is the {\em best statistical strategy} to adapt in order to search
{\em efficiently} for randomly located objects?''

\index{L\'evy!walk}\index{search!efficiency}\index{search!optimality}
To answer this question they introduced a special type of what is
called a {\em L\'evy walk} \cite{ZDK15} in two dimensions and studied
it both by computer simulations and by analytical approximations.
Their model consists of point targets randomly distributed in a plane
and a (point) forager moving with constant speed. If the forager spots
a target within a pre-defined finite vision distance, it moves to the
target directly.  Otherwise the forager chooses a direction at random
with a jump length $\ell$ randomly drawn from a L\'evy stable
distribution $\sim \ell^{-\mu}\:,\:1\le\mu\le3$.  While the forager is
moving it constantly looks out for targets within the given vision
distance. If no target is detected, the forager stops after the given
distance and repeats the process.

Although these rules look simple enough, there are some subtleties
that exemplify the problem of mathematically modeling a biological
foraging problem:

\begin{enumerate}

\item Here we have chosen what is called a {\em cruise forager}, i.e.,
  a forager that senses targets whenever it is moving. In contrast, a
  {\em saltaltory forager} would not sense a target while moving. It
  needs to land close to a target within a given radius of perception
  in order to find it \cite{JPP09}.

\item For a cruise forager a jump is terminated when it hits a target,
  hence this model defines a {\em truncated} L\'evy walk \cite{Sims10}.
\index{L\'evy!walk!truncated}

\item One has to decide whether a forager eliminates targets when it
  finds them or not, i.e., whether it performs {\em destructive} or
  {\em non-destructive} search \cite{Vis99}. As we will see below,
  whether a monkey eats a fruit thus effectively eliminating it, at
  least for a long time, or whether a bee collects nectar from a
  flower that replenishes quickly defines
  mathematically different foraging problems.

\item We have not yet said anything about the {\em density of the targets}.

\item We have deliberately assumed that the targets are {\em immobile},
  which may not always be realistic for a biological foraging problem
  (e.g., marine predators \cite{Sims08,Sims10}).

\item If we ask about the {\em best} strategy to search {\em
    efficiently}, how do we define {\em optimality}?

\end{enumerate}

These few points illustrate the difficulty to relate abtract
mathematical random walk models to biological foraging reality.
Interestingly, the motion generated by these models often sensitively
depends on right such details: In Ref.~\cite{Vis99} foraging
efficiency was defined as the ratio of the number of targets found
divided by the total distance traveled by a forager, see Eq.(3)
therein. Different definitions are possible, depend on the type of
forager and may yield different results \cite{JPP09}. The foraging
efficiency was then computed in Ref.~\cite{Vis99} under variation of
the exponent $\mu$ of the above L\'evy distribution generating the
jump length. The results led to what was coined the {\bf L\'evy
  Flight Hypothesis} (LFH), which we formulate as follows:

L\'evy motion provides an {\em optimal search strategy} for {\em
  sparse, randomly distributed, immobile, revisitable targets in
  unbounded domains.}

\index{foraging}\index{L\'evy!motion}\index{Brownian motion} 

\begin{figure}
\centerline{\includegraphics[height=4cm]{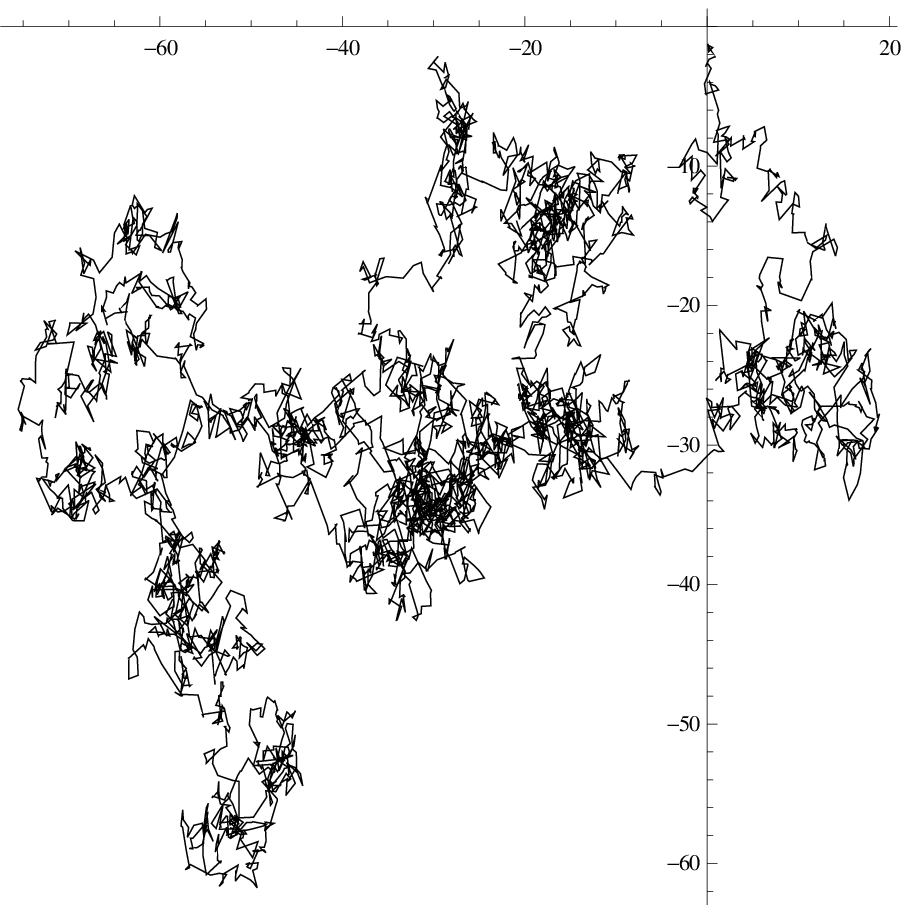}\hspace*{0.5cm}\includegraphics[height=5cm]{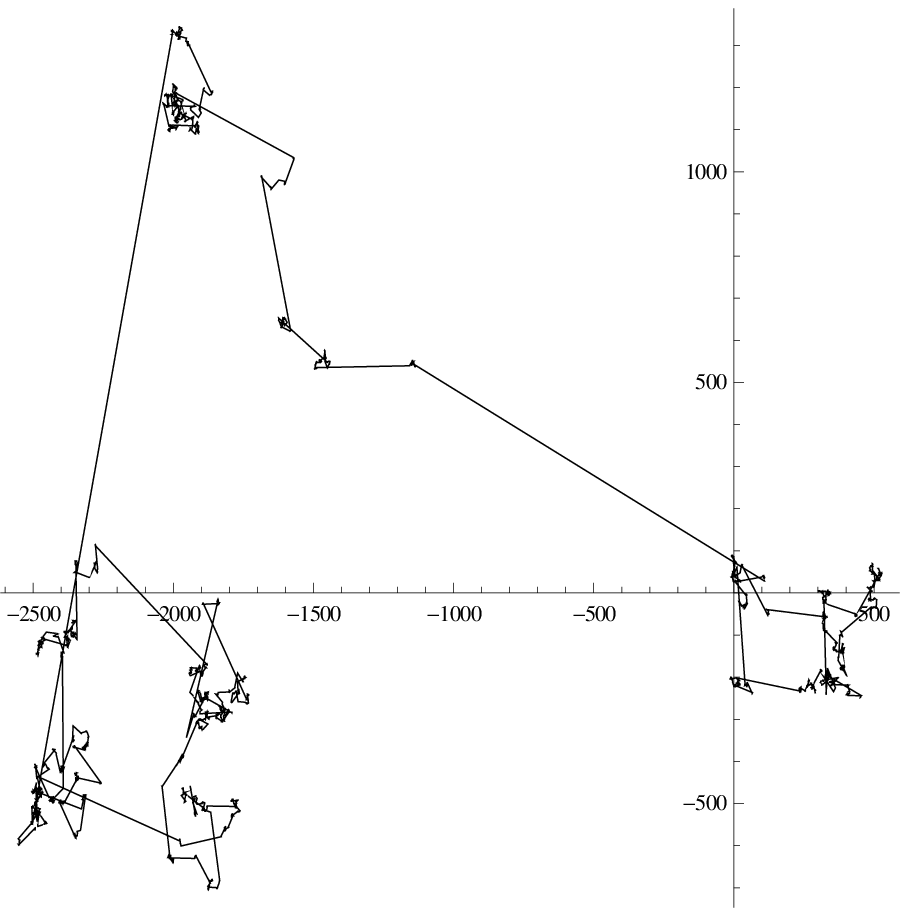}}
\caption{Brownian motion (left) vs.\ L\'evy motion (right) in the
  plane, illustrated by typical trajectories.}
\label{fig:bmlm} 
\end{figure}

Intuitively this result can be understood as follows:
Fig.~\ref{fig:bmlm} (left) displays a typical trajectory of a Brownian
walker. One can see that this dynamics is `more localized' while
L\'evy motion shown in Fig.~\ref{fig:bmlm} (right) depicts clusters
interrupted by long jumps. It thus makes sense that Brownian motion is
better suited to find targets that are densely distributed while
L\'evy motion outperforms Brownian motion when targets are sparse,
since it avoids oversampling due to long jumps. The reason why the
targets need to be revisitable is that the exponent $\mu$ of the
L\'evy distribution depends on whether the search is destructive or
not, cf.\ the third point on the list of foraging conditions
above: For non-destructive foraging $\mu=2$ was found to be optimal
while for destructive foraging $\mu=1$ maximized the foraging
efficiency, which corresponds to the special case of ballistic flights
\cite{ZDK15}. The reason for these different exponents is that
destructive foraging changes the distribution and the density of the
targets thus selecting a different foraging strategy to be optimal.

\section{L\'evy or not L\'evy?}\label{sec:lonl}
  
\subsection{Revisiting L\'evy flights of wandering albatrosses}

Several years passed before the results by Viswanathan et al.\ were
revisited in another {\em Nature} article led by Andrew Edwards
\cite{Edw07}: When analyzing new, larger and more precise data for
foraging albatrosses the old results of Ref.~\cite{Vis96} could not be
recovered, see Fig.~1 in Ref.~\cite{Edw07}. This led the researchers
to reconsider the old albatross data. A correction of these data sets
yielded the result shown in Fig.~\ref{fig:alba} as the red filled
circles: One can see that the L\'evy stable law with an exponent of
$\mu=2$ for the flight times is gone. Instead the data now seems
to be fit best with a gamma distribution.

What happened is explained in Ref.~\cite{Buch08}: For all measurements
the sensors were put onto the feet of the albatrosses when the birds
were sitting on an island, and at this point the measurement process
was started. However, to this time the sensors were dry; and in
Ref.~\cite{Vis96} these times were interpreted as L\'evy flights. The
same applied to the end of a foraging bout when the birds were back on
the island. Subtracting these erroneous time intervals from the data
sets eliminated the L\'evy flights.

However, in Ref.~\cite{HWQ12} yet new albatross data was analyzed, and
the old data from Refs.~\cite{Vis96,Edw07} was again reanalyzed: This
time truncated power laws were used for the analysis, and furthermore
data sets for individual birds were tested instead of pooling together
the data for all birds. In this reference it was concluded that some
individual albatross indeed do perform L\'evy flights while others do
not.
\index{power law!truncated}

\subsection{The L\'evy Flight Paradigm}\label{sec:lfp}

\index{L\'evy!Flight!Paradigm}
The debate about the LFH created a surge of publications testing it
both theoretically and experimentally; see
Refs.~\cite{BLMV11,VLRS11,MCB14,ZDK15} for reviews. But experimentally
it is difficult to verify the mathematical conditions on which the LFH
formulated in Sec.~\ref{sec:tlfh} is based. Often the LFH was thus
interpreted in a much looser sense by ignoring any mathematical
assumptions in terms of what one may call the {\bf L\'evy Flight
  Paradigm} (LFP):

\index{power law}
Look for {\em power laws} in the probability distributions of step
lengths of foraging animals.

\begin{figure}
\centerline{\includegraphics[height=6cm]{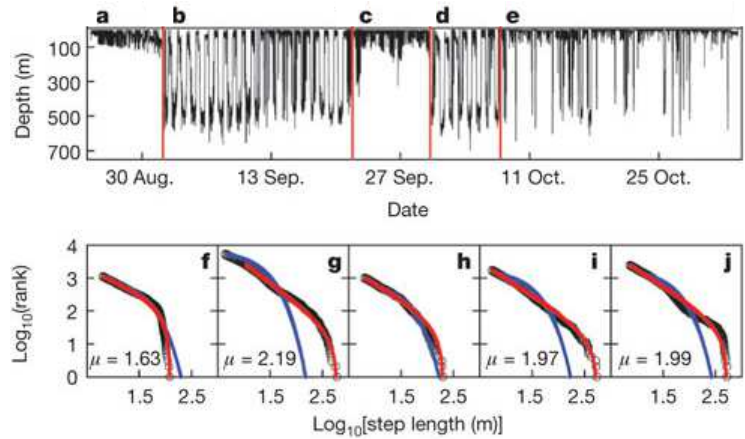}}
\caption{Top: time series of the diving depth of a blue shark. The red
  lines split the data into different sections (a - e), where the
  shark dives deep or the diving depth is more constrained. These
  sections match to the shark being off-shelf or on the shelf,
  respectively. Bottom: double-logarithmic plots of the move step
  length frequency distribution (`rank') as a function of the step
  length, which is the vertical distance moved by the shark per unit
  time, with the notation (f - j) corresponding to the primary data
  shown in sections (a - e).  Black circles correspond to data, red
  lines to fits with truncated power laws of exponent $\mu$, blue
  lines to exponential fits. This figure is reprinted by permission
  from {\em Macmillan Publishers Ltd: Nature Ref.~\cite{Sims10},
    copyright 2010.}}
\label{fig:sims2} 
\end{figure}

We illustrate virtues and pitfalls related to the LFP by data from
Ref.~\cite{Sims10} on the diving depths of free-ranging marine
predators. Impressively, in this work over 12 million movement
displacements were recorded and analyzed for 14 different species. As
an example, Fig.~\ref{fig:sims2} shows results for a blue shark:
Plotted at the bottom are probability distributions of its diving
depths, called move step length frequency distribution, where a step
length is defined as the distance moved by the shark per unit time.
Included are fits to a {\em truncated} power law \index{power
law!truncated} and to an exponential distribution. Since here L\'evy
distributions were used whose longest step lengths were cut off, the
fits do not consist of straight lines but are bent off, in contrast to
Fig.~\ref{fig:alba}. The top of this figure depicts the corresponding
time series from which the data was extracted, split into five
different sections.  Each section is characterized by profoundly
different average diving depths. These different sections correspond
to the shark being in different regions of the ocean, i.e., either
on-shelf or off-shelf. It was argued that on-shelf, where the diving
depth of the shark is very limited, the data can be better fitted with
an exponential distribution (sections f and h) while off-shelf the
data displays power-law behavior with an exponent close to two
(sections g, i and j).  Fig.~\ref{fig:sims2} thus suggests a strong
dependence of the foraging dynamics on the environment in which it
takes place, where the latter defines the food distribution. Related
switching behavior between power law-like L\'{e}vy and exponential
Brownian motion search strategies was reported for microzooplankton,
jellyfish and mussels.

The power law matching to the data in the off-shelf regions was
interpreted in support of the LFH. However, note the periodic
oscillations displayed by the time series at the top of
Fig.~\ref{fig:sims2}. Upon closer inspection they reveal a 24h
day-night cycle: During the night the shark hovers close to the
surface of the sea while over the day it dives for food.  For the move
step length distributions shown in Fig.~\ref{fig:sims2} the data was
averaged over all these periodic oscillations. But the distributions
in sections g, i and j all show a `wiggle' on a finer scale. This
suggests to better fit the data by a superposition of two different
distributions \cite{LICCK12} taking into account that day and night
define two very different phases of motion, instead of using only one
function by averaging over all times. Apart from this, one may argue
that this analysis does not test for the original LFH put forward in
Ref.~\cite{Vis96}. But this requires a bit more knowledge about the
theory of L\'evy motion; we will come back to this point in
Sec.~\ref{sec:lfn}.

\subsection{Two different L\'evy Flight Hypotheses}

Our discussion in the previous sections suggests to distinguish
between {\em two different} LFHs:

\begin{enumerate}

\index{L\'evy!Search Hypothesis} \index{search!strategy!optimal}
\item The first is the `conventional' one that we formulated in
  Sec.~\ref{sec:tlfh}, originally put forward in Ref.~\cite{Vis96}:
  It may now be further specified as the {\bf L\'evy Search
    Hypothesis} (LSH), because it suggests that under certain
  conditions L\'evy flights represent an {\em optimal search
    strategy}. Here optimality needs to be defined rigorously
  mathematically. This can be done in different ways given the
  specific biological situation at hand that one wishes to model
  \cite{JPP09}. Typically optimality within this context aims at
  minimizing the search time for finding targets. The interesting
  biological interpretation of the LSH is that it has been evolved in
  biological organisms as an {\em evolutionary adaptive} strategy that
  maximizes the success for survival. The LSH version of the LFH
  became most popular.

\index{L\'evy!Environmental Hypothesis}\index{emergence}
\item In parallel there is a second type of LFH, which may be called
  the {\bf L\'evy Environmental Hypothesis} (LEH): It suggests that
  L\'evy flights {\em emerge} from the interaction between a forager
  and a food source distribution. The latter may be scale-free thus
  directly inducing the L\'evy flights. This is in sharp contrast to
  the LSH, which suggests that under certain conditions a forager
  performs L\'evy flights irrespective of the actual food source
  distribution. Emergence of novel patterns and dynamics due to the
  interaction of the single parts of a complex system with each other,
  on the other hand, is at the heart of the theory of complex systems.
  The LEH is the hypothesis that to some extent was formulated in
  Ref.~\cite{Vis96}, but it became more popular rather later on
  \cite{RFM03,Sims08,Sims10}.

\end{enumerate}

\index{L\'evy!Flight!Paradigm}
Both the LSH and the LEH are bound together by what we called the
L\'evy Flight Paradigm (LFP) in Sec.~\ref{sec:lfp}. The LFP extracts
the formal essence from both these different hypotheses by proposing
to look for power laws in the probability distributions of foraging
dynamics by ignoring any conditions of validity of these two
hypotheses. Consequently, in contrast to the LSH and LEH the
mathematical, physical and biological origin and meaning of power laws
obtained by following the LFP is typically not clear. On the other
hand, the LFP motivated to take a fresh look at foraging data sets by
not only testing for exponential distributions. It widened the scope
by emphasizing that one should also check for power laws in animal
movement data.

\subsection{Intermittent search strategies as an alternative to L\'evy motion}\label{sec:int}

\begin{figure}
\centerline{\includegraphics[height=2.8cm]{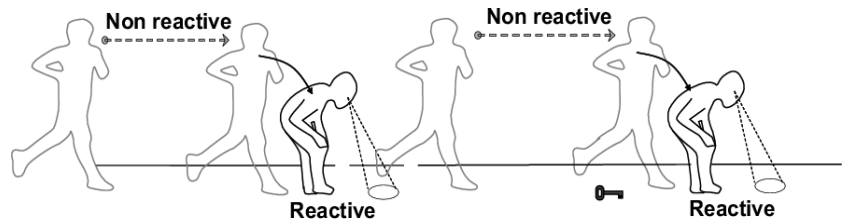}}
\caption{Illustration of an intermittent search strategy: A human
  searcher looks for a target (key) by alternating between two
  different modes of motion. During fast, ballistic relocation phases
  the searcher is not able to detect any target (non reactive). These
  phases are interrupted by slow phases of Brownian motion during
  which a searcher is able to detect a target (reactive)
  \cite{BLMV11}.}
\label{fig:interm} 
\end{figure}

\index{intermittency}\index{search!strategy!intermittent}
Simple random walks as introduced in Section 2.1 represent examples
of {\em unimodal} types of motion if the random step lengths are
sampled from only one specific distribution. For example, choosing a
Gaussian distribution we obtain Brownian motion while a L\'evy-stable
distribution produces L\'evy flights. Combining two different types of
motion like Brownian and L\'evy yields {\em bimodal motion}. A simple
example is shown in Fig.~\ref{fig:interm}: Imagine you have lost your
keys at home, but you have a vague idea where to find them. Hence, you
are running straightforwardly to the location where you expect them to
be. This may be modeled as a {\em ballistic flight} during which you
quickly relocate, say, from the kitchen to the study room.  However,
when you arrive in your study room you should switch to a different
type of motion, which is suitably adapted to locally search the
environment. For this mode you may choose, e.g., Brownian motion.  The
resulting dynamics is called {\em intermittent} \cite{BLMV11}: It
consists of two different phases of motion mixed randomly, which in
our example are ballistic relocation events and local Brownian motion.

\index{search}
This type of motion can be exploited to search efficiently in the
following way: You may not bother to look for your keys while you are
walking from the kitchen to the study room.  You are more interested
to get from point A to point B as quickly as possible, and while doing
so your search mode is switched off. This is called a {\em non
  reactive} phase in Fig.~\ref{fig:interm}. But as you expect the keys
to be in your study room, while switching to Brownian motion therein
you simultaneously switch on your scanning abilities.  This defines
your local search mode called {\em reactive} in
Fig.~\ref{fig:interm}. Correspondingly, for aninmals one may imagine
that during a fast relocation event, or flight, they are unable to
detect any targets while their sensory mechanisms become active during
slow local search. This is close to what was called a saltaltory
forager in Sec.~\ref{sec:tlfh}, but this forager did not feature
any local search dynamics.

Intermittent search dynamics can be modeled by writing down a set of
two coupled equations, one that generates ballistic flights and
another one that yields Brownian motion. The coupling captures the
switching between both modes. One furthermore needs to model that
search is only performed during the Brownian motion mode. By analyzing
a respective ballistic-Brownian system of equations it was found that
this dynamics yields a minimum of a suitably defined search time under
parameter variation if a target is {\em non-revisitable}, i.e., it is
destroyed once it is found. Note that for targets that are
non-replenishing the L\'evy walks of Ref.~\cite{Vis99} did not yield
any non-trivial optimization of the search time.  Instead, they
converged to pure ballistic flights as being optimal. The LSH, in
turn, only applies to revisitable, i.e., replenishing targets. Hence
intermittent motion poses no contradiction.  A popular account of this
result was given by Michael Shlesinger in his Nature article `How to
hunt a submarine?'  \cite{Shles06}.

\subsection{Theory of L\'evy flights in a nutshell}\label{sec:lfn}

\index{L\'evy!motion}\index{L\'evy!Flight} We now briefly elaborate on
the theory of L\'evy motion. This section may be skipped by a reader
who is not so interested in theoretical foundations. We recommend
Ref.~\cite{SZK93} for an outline of this topic from a physics point of
view and Chap.~5 in Ref.~\cite{KRS08} for a more mathematical
introduction. We start from the simple random walk on the line
introduced in Chap.~2 of this book, \index{random walk} \be
x_{n+1}=x_n+\ell_n\quad , \label{eq:lm} \ee where $x_n$ is the
position of a random walker at discrete time $n\in\mathbb{N}$ moving
in one dimension, and $\ell_n=x_{n+1}-x_n$ defines the jump of length
$|\ell_n|$ between two positions. In Chap.~2 the special case of
constant jump length $|\ell_n|=\ell$ was considered, where the sign of
the jump was randomly determined by tossing a coin with, say, plus for
heads and minus for tails. The coin was furthermore supposed to be
{\em fair} in the sense of yielding equal probabilities for heads and tails.
This simple random walk can be generalized by considering a bigger
variety of jumps. Mathematically this is modeled by drawing the random
variable $\ell_n$ from some more general probability distribution than
featuring only probability one half for each of two outcomes. For
example, instead we could draw $\ell_n$ at each time step $n$ randomly
from a uniform distribution, where each jump between $-L$ and $L$ is
equally possible given by the probability density $\rho(\ell_n)=1/(2L)
\: , \: -L\le\ell_n\le L$ and zero otherwise.  Alternatively, we could
allow arbitrarily large jumps by drawing $\ell_n$ from an unbounded
Gaussian distribution, see Eq.(2.10) in Chap.~2 (by replacing $x$
therein with $\ell_n$ and setting $t$ constant). For both generalized
random walks Eq.~(\ref{eq:lm}) would still reproduce in the long time
limit the fundamental diffusive properties Eq.~(4) discussed in
Chap.~2, i.e., the linear growth in time of the mean square
displacement, and Eq.~(2.10) in Chap.~2, the Gaussian probability
distribution for the position $x_n$ of a walker at time step $n$. This
follows mathematically from the conventional central limit
theorem. \index{central limit theorem}

We now further generalize the random walk Eq.~(\ref{eq:lm}) in a more
non-trivial way by randomly drawing $\ell_n$ from a {\em L\'evy
  $\alpha$-stable distribution} \cite{KRS08},
\index{L\'evy!stable distribution}
\be
\rho(\ell_n)\sim |\ell_n|^{-1-\alpha}\: (|\ell_n|\gg1)\:,\:0<\alpha<2 \label{eq:lsd} \:,
\ee
characterized by power law tails in the limit of large $|\ell_n|$. This
functional form is in sharp contrast to the exponential tails of
Gaussian distributions and has important consequences, as it violates
one of the assumptions on which the conventional central limit theorem
rests.  However, for the range of exponents $\alpha$ stated above
it can be shown that these distributions obey a {\em generalized
  central limit theorem}: \index{central limit theorem!generalized}
The proof employs the fact
that these distributions are {\em stable}, in the sense that a linear
combination of two random variables sampled independently from the
same distribution reproduces the very same distribution, up to some
scale factors \cite{SZK93}. This in turn implies that L\'evy stable
distributions are {\em scale invariant} and thus {\em self-similar}.
Physically one speaks of $\ell_n$ sampled independently and
identically distributed from Eq.~(\ref{eq:lsd}) as {\em white L\'evy
  noise}. \index{L\'evy!noise}
As by definition there are no correlations between the
random variables $\ell_n$ the stochastic process generated by
Eq.~(\ref{eq:lm}) is {\em memoryless}, meaning at time step $(n+1)$ the
particle has no memory where it came from at any previous time step
$n$. In mathematics this is called a {\em Markov processes}, and
L\'evy flights belong to this important class of stochastic processes.
\index{Markov process}

What we presented here is only a very rough, mathematically rather
imprecise outline of how to define an $\alpha$-stable L\'evy process
generating L\'evy flights. Especially, the function in
Eq.~(\ref{eq:lsd}) is not defined for small $\ell_n$, as the given
power law diverges for $\ell_n\to0$. A rigorous definition of L\'evy
stable distributions is obtained by using the characteristic function
of this process, i.e., the Fourier transform of its probability
distribution, which is well-defined analytically. The full probability
distribution can then be generated from it \cite{SZK93,KRS08}. For
$\alpha=2$ this approach reproduces Gaussian distributions, hence
L\'evy dynamics suitably generalizes Brownian motion
\cite{SZK93,KRS08}.

Another important property of L\'evy stable distributions is that the mean
squared flight length of a L\'evy walker does not exist,
\be
\langle \ell_n^2 \rangle=\int_{-\infty}^{\infty} \; d \ell_n \; \rho(\ell_n) \ell_n^2=
\infty\: .
\ee
The above equation defines what is called the second moment of the
probability distribution $\rho(\ell_n)$. Higher moments are defined
analogously by $\langle \ell^k \rangle\:,\:k\in\mathbb{N}$, and for
L\'evy distributions they are also infinite.  This means that in
contrast to simple random walks generating Brownian motion, see again
Chap.~2, L\'evy motion does not have any characteristic length
scale. However, since moments are rather easily obtained from
experimental data this poses a problem to L\'evy flights as a viable
physical model to be validated by experiments.

\index{L\'evy!walk}
This problem can be solved by using the very related concept of {\em
  L\'evy walks} \cite{ZDK15}: These are random walks where again jumps
are drawn randomly from the L\'evy stable distribution
Eq.~(\ref{eq:lsd}). But as a penalty for long jumps the walker spends
a time $t_n$ proportional to the length of the jump to complete it,
$t_n=v\ell_n$, where the proportionality factor $v$, typically chosen
as $|v|=const.$, defines the velocity of the L\'evy walker. This
implies that both jump lengths $\ell_n$ and flight times $t_n$ are
distributed according to the same power law. In contrast, for the
L\'evy flights introduced above a walker makes a jump of length
$|\ell_n|$ during an integer time step of duration $\Delta n=1$, which
implies that contrary to a L\'evy walker a L\'evy flyer can jump
instantaneously over arbitrarily long distances with arbitrarily large
velocities.

\index{random walk!continuous time}
L\'evy walks belong to the broad and important class of {\em
  continuous time random walks} \cite{MeKl00,KRS08,KlSo11}, which
further generalize ordinary random walks by allowing a walker to move
by non-integer time steps. We do not discuss all the similarities and
differences between L\'evy walks and L\'evy flights, see
Ref.~\cite{ZDK15} for details, but instead highlight only one
important fact: While for L\'evy flights the mean square displacement
$\langle x^2\rangle$, see Eq.(1) in Chap.~2, does not exist, which
follows from our discussion above, for L\'evy walks it does. This is
due to the finite velocities, which truncate the power law tails in
the probability distributions for the positions of a L\'evy walker.
However, in contrast to Brownian motion where it grows linearly in
time as shown in Chap.~2, see Eq.(2), for L\'evy walks it grows
faster than linear,
\be
\langle x^2\rangle\sim t^{\gamma}\:(t\to\infty)\:,
\ee
\index{diffusion!anomalous}\index{superdiffusion}\index{subdiffusion}
with $\gamma>1$. If $\gamma\neq1$ one speaks of {\em anomalous
  diffusion} \cite{MeKl00,KRS08}. The case $\gamma>1$ is called {\em
  superdiffusion}, since a particle diffuses faster than Brownian
motion, correspondingly $\gamma<1$ refers to {\em subdiffusion}. There
is a wealth of different stochastic models exhibiting anomalous
diffusion, and while superdiffusion appears to be more common among
foraging biological organisms than subdiffusion the whole spectrum of anomalous
diffusion is found in a variety of different processes in the natural
sciences, and even in the human world \cite{MeKl00,MeKl04,KRS08}.

\index{power law!truncated}
Often the difference between L\'evy walks and flights is not quite
appreciated in the experimental literature, see, e.g.,
Fig.~\ref{fig:sims2}, where move step length frequency distributions
were plotted. By definition a move step length $x$ per unit time
corresponds to what we defined as a jump length $\ell_n$ by
Eq.~(\ref{eq:lm}) above, $x=\ell_n$. Hence, a truncated power law fit
$\sim x^{-\mu}$ to the distributions plotted in Fig.~\ref{fig:sims2}
corresponds to a fit with a truncated form of the jump length
distribution Eq.~(\ref{eq:lsd}) with exponent $\mu=1+\alpha$ testing
for truncated L\'evy flights \cite{ZDK15}. The truncation cures the
problem of infinite moments exhibited by random walks based on
ordinary L\'evy flights mentioned above. However, this analysis does
not test the LFH put forward in Ref~\cite{Vis99}, which was derived
from L\'evy walks.  But checking for L\'evy walks requires an entirely
different data analysis \cite{HaBa12,ZDK15}.

\section{Beyond the L\'evy Flight Hypothesis: foraging
  bumblebees}\label{sec:bee}
  
\index{bumblebee}
The LFH and its variants illustrated the problem to which extent
biologically relevant search strategies may be identified by
mathematical modeling. What we then formulated as the LFP in
Sec.~\ref{sec:lfp} motivated to generally look for power laws in
the probability distributions of step lengths of foraging animals.
Inspired by the long debate about the different functional forms of
move step lengths probability distributions, and by further diluting
the LFP, an even weaker guiding principle would be to assume that the
foraging dynamics of biological organisms can be understood by
analyzing such probability distributions alone. In the following we
discuss an experiment, and its theoretical analysis, which illustrate
that one may miss crucial information by studying only probability
distributions. In that respect, this last section provides a look
beyond the LFH that focuses on such distributions.

\subsection{Bumblebees foraging under predation risk}

In Refs.~\cite{Ings08} Thomas Ings and Lars Chittka reported a
laboratory experiment in which environmental foraging conditions were
varied in a fully controlled manner. The question they addressed with
this experiment was whether changes of environmental conditions, in
this case exposing bumblebees to predation threat or not, led to
changes in their foraging dynamics. This question was answered by a
statistical analysis of the bumblebee flights recorded in this
experiment on both spatial and temporal scales \cite{LICCK12}.

\begin{figure}
\centerline{(a)\includegraphics[height=4cm]{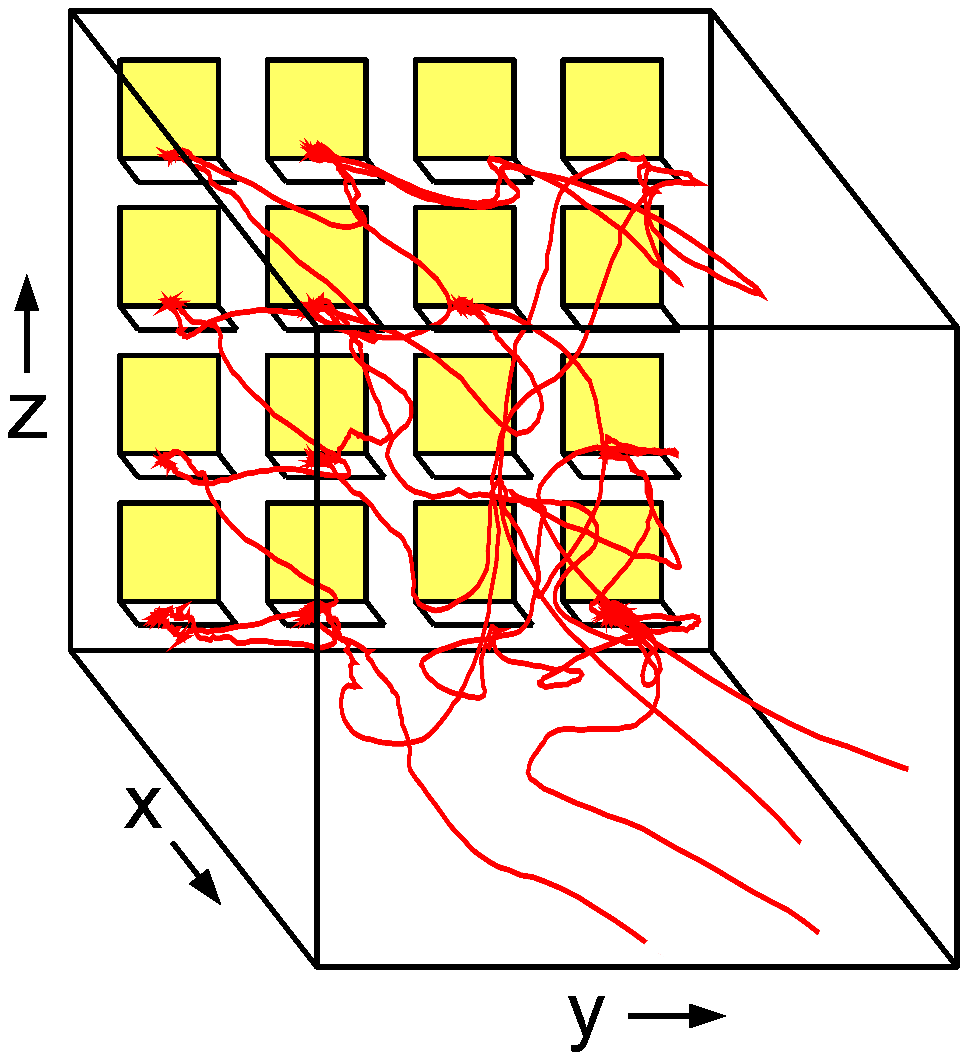}\hspace*{0.5cm}(b)\includegraphics[height=4cm]{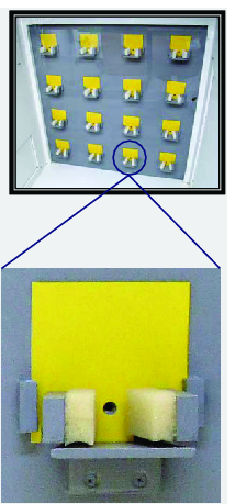}\hspace*{0.5cm}(c)\includegraphics[height=4cm]{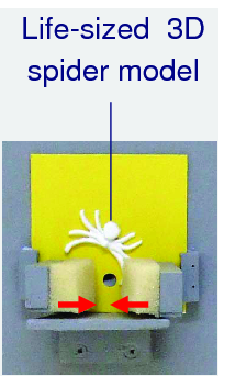}}
\caption{Illustration of a laboratory experiment investigating the
  dynamics of bumblebees foraging under predation risk: (a) Sketch of
  the cubic foraging arena together with part of the flight trajectory
  of a single bumblebee. The bumblebees forage on a grid of artificial
  flowers on one side of the box.  While being on the landing
  platforms, they have access to nectar.  All flowers can be equipped
  with spider models and trapping mechanisms simulating predation
  attempts as shown in (b), (c) \cite{Ings08,LICCK12}.}
\label{fig:beeexp} 
\end{figure}

\index{foraging} The experiment is sketched in Fig.~\ref{fig:beeexp}:
Bumblebees ({\em Bombus terrestris}) were flying in a cubic arena of
$\approx$ 75cm side length by foraging on a 4$\times$4 vertical grid
of artificial yellow flowers on one wall. The 3D flight trajectories
of 30 bumblebees, tested sequentially and individually, were tracked
by two high frame rate cameras. On the landing platform of each flower
nectar was given to the bumblebees and replenished after consumption.
To analyze differences in the foraging behavior of the bumblebees
under threat of predation, artificial spiders were introduced. The
experiment was staged into several different phases of which, however,
only the following three are relevant to our analysis:
\begin{enumerate}
\item spider-free foraging
\item foraging under predation risk
\item a memory test one day later
\end{enumerate}
Before and directly after stage 2 the bumblebees were trained to
forage in the presence of artificial spiders, which were randomly
placed on 25\% of the flowers. A spider was emulated by a spider model
on the flower and a trapping mechanism, which briefly held the
bumblebee to simulate a predation attempt. In stages 2 and 3 the
spider models were present but the traps were inactive in order to
analyze the influence of previous experience with predation risk on
the bumblebees' flight dynamics; see Ref~\cite{Ings08} for full
details of the experimental setup and staging.

It is important to observe that neither the LSH nor the LEH can be
tested by this experiment, as the flight arena is too small: The
bumblebees always sense the walls and may adjust their flight behavior
accordingly. However, there is a cross-link to the LEH in that this
experiment studies the interaction of a forager with the environment,
and its consequences for the dynamics of the forager, in a very
controlled way. The weaker guiding principle derived from the LFP that
we discussed above furthermore suggests that the main information to
understand the foraging dynamics may be contained in the probability
distributions of flight step lengths only. On this basis one may
naively expect to see different step lengths probability distributions
emerging by changing the environmental conditions, which here is the
predation risk. 

\subsection{Velocity distributions vs.\ velocity correlations:
  experimental results}

\begin{figure}
    \centerline{\includegraphics[height=7cm,angle=-90]{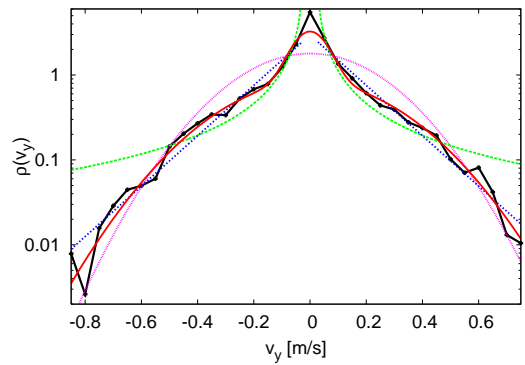}}
    \caption{ \label{fig:beepdf} Semi-logarithmic plot of the
      distribution of velocities $v_y$ parallel to the $y$-axis in
      Fig.~\ref{fig:beeexp}(a) (black crosses) for a single bumblebee
      in the spider-free stage 1. The different lines represent
      maximum likelihood fits with a Gaussian mixture (red line),
      exponential (blue dotted), power law (green dashed), and single
      Gaussian distribution (violet dotted) \cite{LICCK12}.}
\end{figure}

Figure~\ref{fig:beepdf} shows a typical probability distribution of
the horizontal velocities parallel to the flower wall (cf.\ the
y-direction in Fig.~\ref{fig:beeexp}(a)) for a single bumblebee. This
distribution is in analogy to the move step length frequency
distributions of the shark shown in Fig.~\ref{fig:sims2}, which also
represent velocity distributions if the depicted step lengths are
divided by the corresponding constant time intervals of their
measurements as discussed in Sec.~\ref{sec:lfn}. The distribution of
bumblebee flights per unit time is characterized by a peak at low
velocities. Only a power law and a Gaussian distribution can
immediately be ruled out by visual inspection as matching functional
forms. However, a mixture of two Gaussian distributions and an
exponential function appear to be equally possible. Maximum likelihood
fits supplemented by respective information criteria yielded the
former as the most likely functional form matching the data. This
result can be understood biologically as representing two different
flight modes near a flower versus far away from it, which is confirmed
by spatially separated data analysis \cite{LICCK12}. That the
bumblebee switches to a specific distribution of lower velocities when
approaching a flower reflects a spatially adapted flight mode to
accessing the food sources. As a result, here we encounter another
version of intermittent motion: In contrast to the temporal switching
between different flight modes discussed in Sec.~\ref{sec:int} this
one is due to switching in different regions of space.
 
Surprisingly, when extracting the velocity distributions of single
bumblebees at the three different stages of the experiment and
comparing their best fits with each other, qualitatively and
quantitatively {\em no differences} could be found in these
distributions between the spider-free stage and the stages where
artificial spider models were present \cite{LICCK12}. This means that
the bumblebees fly with the very same statistical distribution of
velocities irrespective of whether predators are present or not. The
answer about possible changes in the bumblebee flights due to changes
in the environmental conditions is thus not given by analyzing the
probability distributions of move step lengths, as one may infer from
our diluted LFP guiding principle. We will now see that it is provided
by examining the correlations of horizontal velocities $v_y(t)$
parallel to the wall for all bumblebee flights. They can be measured
by the {\em velocity autocorrelation function}
\index{velocity autocorrelation function}\index{covariance}\index{memory}
\index{correlation}
\be
v_y^{ac}(\tau)=\frac{\left< (v_y(t)-\mu)(v_y(t+\tau)-\mu) \right>}{\sigma^2}
\label{eq:vacf}\:.
\ee
Here $\mu $ and $\sigma^2$ denote the mean and the variance of the
corresponding velocity distribution of $v_y$, respectively, and the
angular brackets define an average over all bumblebees and over
time. This quantity is a special case of what is called a {\em
  covariance} in statistics. Note that velocity correlations are
intimately related to the mean square displacement introduced in
Chap.~2 of this book: While the above equation defines velocity
correlations that are normalized by subtracting the mean and dividing
by the variance, unnormalized velocity correlations emerge
straightforwardly from the right hand side of Eq.~(2.1) in Chap.~2
by rewriting it as products of velocities. This yields the {\em
  (Taylor-)Green-Kubo formula} expressing the mean square displacement
exactly in terms of velocity correlations \cite{Kla06}. Note that the
velocity autocorrelation function is defined by an average over the
product between the initial velocity at time $\tau=0$ and the velocity
at time lag $\tau$ along a trajectory: By definition it is maximal and
normalized to one at $\tau=0$, because the initial velocity is
maximally correlated with itself. It will decay to zero if on average
all velocities at time $\tau$ are randomly distributed with respect to
the initial velocities. Physically this quantity thus measures the
{\em correlation decay} in the dynamics over time $\tau$ by giving an
indication to which extent a dynamics loses memory. For example, for a
simple random walk as defined in Chap.~2 and by Eq.~(\ref{eq:lm})
in our section the velocity correlations would immediately jump to
zero from $\tau=0$ to $\tau\neq0$, which reflects that these random
walks are completely memory-free. This property was used in
Chap.~2 to derive Eq.~(2.2) from Eq.~(2.1) by canceling all
cross-correlation terms.

\begin{figure}
        \centerline{(a)\includegraphics[height=5cm,angle=-90]{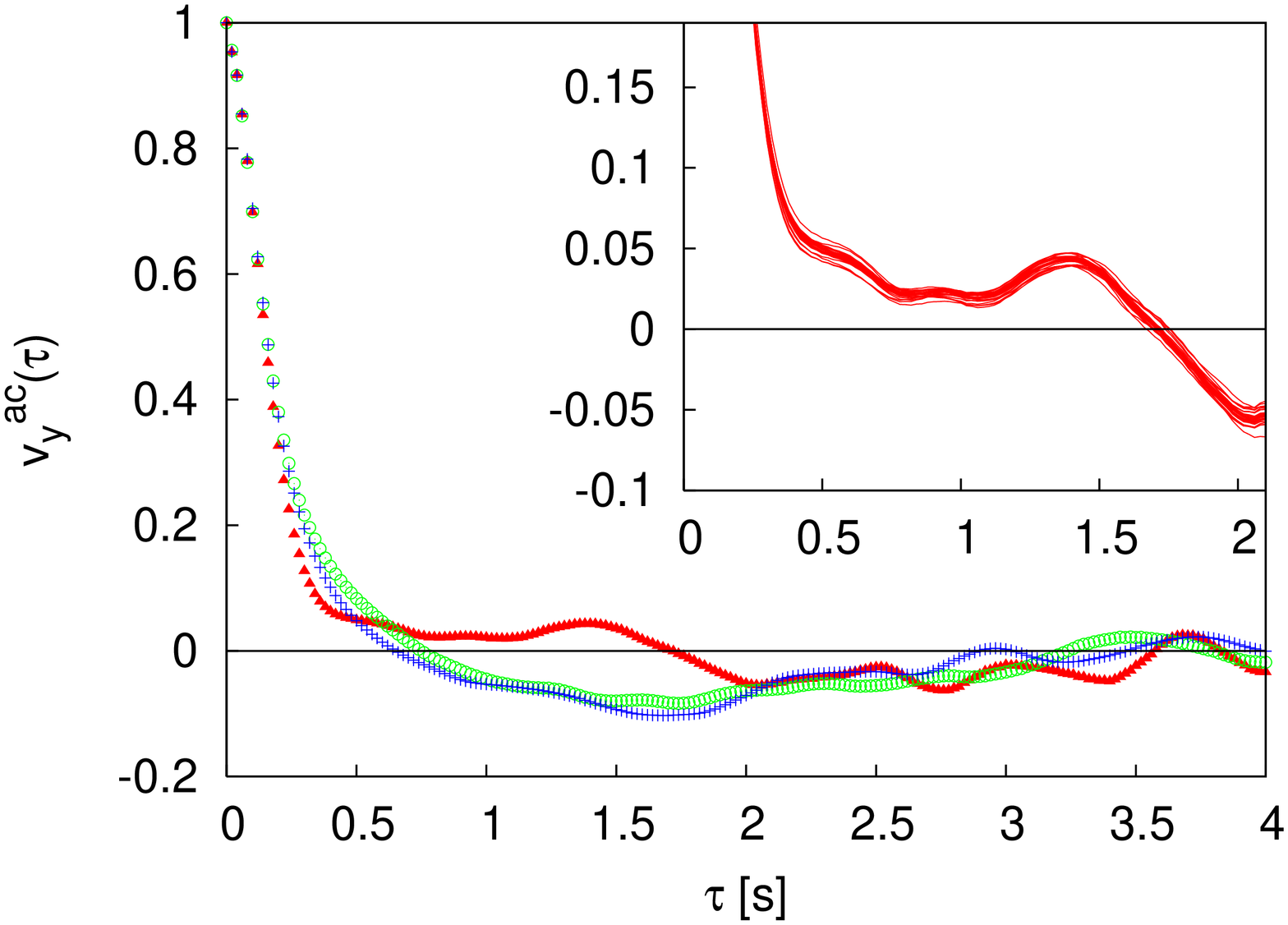}\hspace*{0.2cm}(b)\includegraphics[height=5cm,angle=-90]{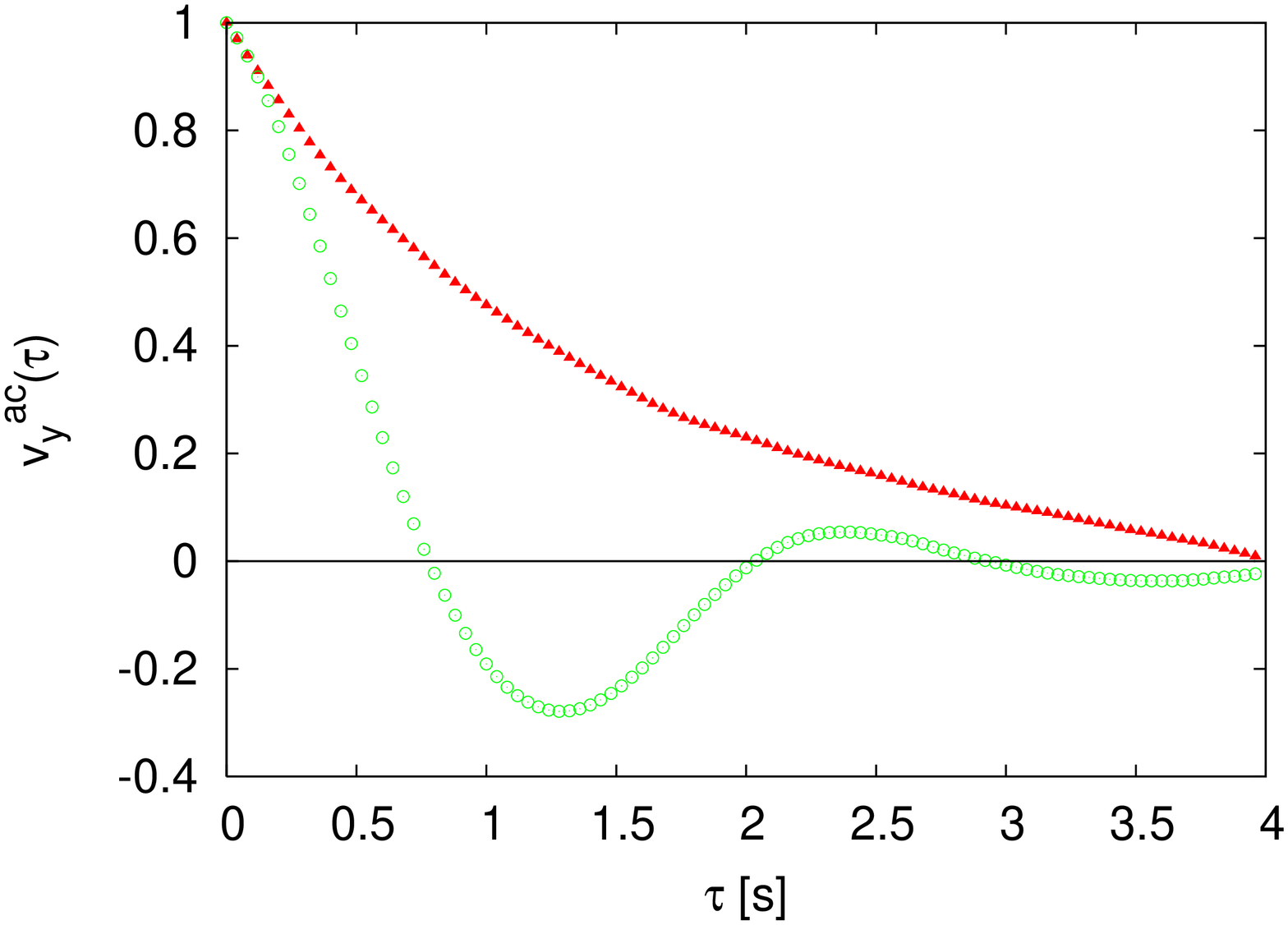}}
        \caption{ \label{fig:beecorr} Velocity autocorrelation
          function Eq.~(\ref{eq:vacf}) for bumblebee velocities $v_y$
          parallel to the wall at three different stages of the
          experiment shown in Fig.~\ref{fig:beeexp}: (a) Experimental
          results for stage 1 without spiders (red), 2 under predation
          threat (green), and 3 under threat a day after the last
          encounter with the spiders (blue). The data show the effect
          of the presence of spiders on the bumblebee flights. The
          inset presents the resampled autocorrelation for the
          spider-free stage in the region where the correlation
          differs from the stages with spider models, which confirms
          that the positive autocorrelations are not a numerical
          artifact. (b) Theoretical results for the same quantity
          obtained from numerically solving the Langevin
          equation~(\ref{eq:ule}) by switching off (red triangles,
          upper line) / on (green circles, lower line) a repulsive
          force modeling the interaction of a bumblebee with a spider.
          These results qualitatively reproduce the experimental
          findings in (a).} \index{velocity autocorrelation function}
\end{figure}

Figure~\ref{fig:beecorr}(a) shows the bumblebee velocity
autocorrelations defined by Eq.~(\ref{eq:vacf}) for all three stages
of the experiment. While for the spider-free stage the correlations
remain positive for rather long times, in the presence of spiders they
quickly become negative. This means that the velocities are on average
anti-parallel to each other, or anti-correlated. In terms of flights,
when predators are not present the bumblebees thus fly on average more
often in the same direction for short times while in the presence of
predators on average they often reverse their flight directions for
shorter times. This result can be biologically understood as
reflecting a more careful search under predation threat: When no
predators are present, the bumblebees forage with more or less direct
flights from flower to flower. However, under threat the bumblebees
change their direction more often in their search for food sources,
rejecting flowers with spiders.  Mathematically this means that the
{\em distributions} of velocities remain the same, irrespective of
whether predators are present or not, while the {\em topology}, i.e.,
the shape of the bumblebee trajectories changes profoundly being on
average more `curved'.

\index{Langevin equation}
In order to theoretically reproduce these changes we model the
dynamics of $v_y$ by a {\em Langevin equation} \cite{Reif}. It may be
called Newton's Law of stochastic physics, as it is based on Newton's
Second Law: $F=m\cdot a$, where $m$ is the mass of a tracer particle
in a fluid moving with acceleration $a=d^2x/dt^2$ at position $x(t)$
(for sake of simplicity we restrict ourselves to one dimension). To
model the interaction of the tracer particle with the surrounding
fluid, the force $F$ on the left hand side is written as a sum of two
different forces, $F=F_S+F_b$: a friction term $F_S=-\eta v=-\eta
\;dx/dt$ with Stokes friction coefficient $\eta$, which models the
damping by the surrounding fluid; and another term $F_b$ that mimicks
the microscopic collisions of the tracer particle with the surrounding
fluid particles, which are supposed to be much smaller than the tracer
particle. The latter interaction is modeled by a stochastic force
$\xi(t)$ of the same type as we have described in Sec.~\ref{sec:lfn}
for which here one takes Gaussian white noise. Interestingly, the
stochastic Langevin equation can be derived from first principles
starting from Newton's microscopic equations of motion for the full
deterministic dynamical system of a tracer particle interacting with a
fluid consisting of many particles \cite{Kla06}.

At first view it may look strange to apply such an equation for
modeling the motion of a biological organism. However, for a bumblebee
the force terms may simply be reinterpreted: While the friction term
still models the loss of velocity due to the surrounding air during a
flight, the stochastic force term now mimicks both the force actively
exerted by the bumblebee to perform a flight and the randomness of
these flights due to the surrounding air, and to sudden changes of
direction by the bumblebee itself. In addition, for our experiment we
need to model the interaction with predators by a third force
term. This leads to Eq.~(20) stated in Chap.~2, which for
bumblebee velocities $v_y$ we rewrite as
\be	
\frac{dv_y(t)}{dt} = - \eta v_y(t) - \frac{d U(y(t))}{d y} + \xi(t)\:. \label{eq:ule}
\ee
Here we have combined the mass $m$ with the other terms on the right
hand side. The term $F_i=- d U(y(t))/dy$ with potential $U$ mimics an
interaction between bumblebee and spider, which can be switched on or
off depending on whether a spider is present or not. Data analysis
shows that this force is strongly repulsive \cite{LICCK12}. Computing
the velocity autocorrelation function Eq.~(\ref{eq:vacf}) by solving
the above equation numerically for a suitable choice of a repulsive
force qualitatively reproduces a change from positive to negative
correlations when switching on the repulsive force, see
Fig.~\ref{fig:beecorr}(b).

These results demonstrate that velocity correlations can contain
crucial information for understanding foraging dynamics, here in the
form of highly non-trivial correlation decay {\em emerging} from the
interaction of a forager with predators. This experiment could not
test the LSH, as the mathematical assumptions on its validity were not
fulfilled. However, conceptually these results are in line with the
idea underlying the LEH: Theoretically the interaction between forager
and environment was modeled by a repulsive force, to be switched on in
the presence of predators, which qualitatively reproduced the
experimental results. Together with the spatially intermittent
dynamics when approaching the food sources as discussed before, these
findings illustrate a complex spatio-temporal adjustment of the
bumblebees both to the presence of food sources and predators. This is
in sharp contrast to the scale-free dynamics singled out by
the LFH.

Of course, modeling bumblebee flights by a Langevin equation like
Eq.~(\ref{eq:ule}) ignores many fine details.  A more sophisticated
model that reproduces bumblebee flights far away from the flowers more
appropriately has been constructed in Ref.~\cite{LCK13} based on the
same data as discussed above. 

\section{L\'evy flights embedded in Movement  Ecology}\label{sec:summ}
  
\index{movement ecology}\index{search}\index{emergence}
\index{L\'evy!Flight!Hypothesis}\index{L\'evy!Flight!Paradigm}
\index{L\'evy!Environmental Hypothesis}\index{L\'evy!Search Hypothesis}
The main theme of our chapter was the question posed to the end of the
introduction: {\em Can search for food by biological organisms be
  understood by mathematical modeling?} While about a century ago this
question was answered by Karl Pearson in terms of simple random walks
yielding Brownian motion, about two decades ago the LFH gave a
different answer by proposing L\'evy motion to be optimal for foraging
success, under certain conditions. Discussing experimental results
testing it, we arrived at a finer distinction between two different
types of LFHs: The LSH captured the essence of the original LFH by
stating that under certain conditions L\'evy flights represent an
optimal search strategy for finding targets. In contrast the LEH
stipulates that L\'evy flights may emerge from the interaction between
a forager and possibly scale-free food source distributions. A weaker
version of these different hypotheses we coined the LFP, which
suggests to look for power laws in the probability distributions of
move step lengths of foraging organisms. An even weaker guiding
principle derived from it is to assume that the foraging dynamics of
biological organisms can generally be understood by analyzing step
length probability distributions alone. We thus have a hierarchy of
different LFHs that have all been tested in the literature, in one way
or the other.

By elaborating on experimental results, exemplified by selected
publications, we outlined a number of problems when testing the
different LFHs: miscommunication between theorists and
experimentalists leading to incorrect data analysis; the difficulties
to mathematically model a specific foraging situation by giving proper
credit to all relevant biological details; and problems with an
adequate statistical data analysis that really tests for the theory by
which it was motivated. We highlighted that there are alternative
stochastic processes, such as intermittent search strategies, that may
outperform L\'evy strategies under certain conditions, or at least
lead to similar results, such that it may be hard to clearly
distinguish them from L\'evy motion. We also discussed an experiment
on foraging bumblebees, which showed that relevant information to
understand a biological foraging process may not always be contained
in the probability distributions that are at the heart of all versions
of the LFH. These experimental results suggested that biological
organisms may rather perform a complex spatio-temporal adjustment to
optimize their search for food sources, which results in different
dynamics on different spatio-temporal scales. This is at variance to
L\'evy motion, which by definition is scale-free.

\begin{figure}
\centerline{\includegraphics[width=8cm]{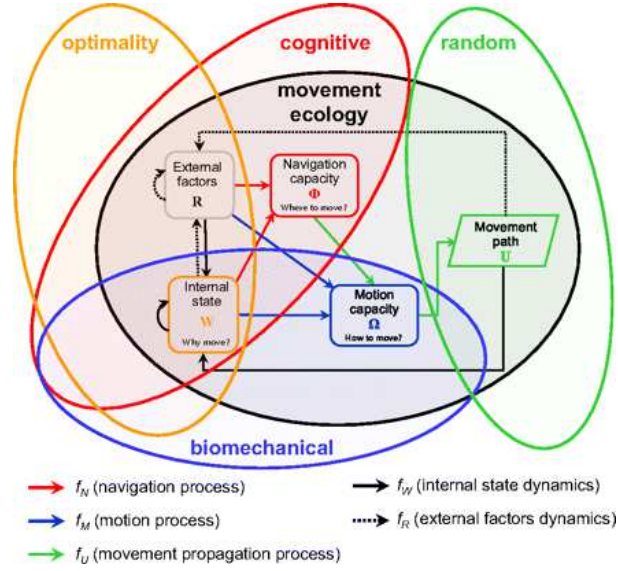}}
\caption{ \label{fig:mep} Sketch of the {\em Movement Ecology
    Paradigm}: It cross-links four other existing paradigms
  representing different scientific disciplines, which describe
  specific aspects of the movements of biological organisms. The aim
  is to mathematically model the dynamics emerging from the interplay
  between these different fields by an equation like
  Eq.~(\ref{eq:mepssa}); from \cite{Nath08}, \index{movement ecology}
  {\em copyright (2008) National Academy of Sciences, U.S.A.}}
 \end{figure}

However, these results are well in line with another, more general
approach to understand the movements of biological organisms, called
the {\em Movement Ecology Paradigm} \cite{Nath08}: This theory aims at
more properly embedding the movements of biological organisms into
their biological context as shown in Fig.~\ref{fig:mep}. In this
figure, the theory centered around the LFH is rather represented by
the region labeled `random', which focuses on analyzing movement paths
only. However, movement paths of organisms cannot properly be
understood without embedding them into their biological context: They
are to quite some extent determined by the cognitive abilities of the
organisms and their biomechanical abilities, see the respective two
further regions in this diagram. Indeed, only on this basis the
question about optimality may be asked, cf.\ the fourth region in this
diagram, which here is rather understood in a biological sense than as
purely mathematical efficiency. Physicists and mathematicians are used
to think of diffusive spreading, which underlies foraging, primarily
in terms of moving point particles; however, living biological
organisms are not point particles but interact with the surrounding
world in a very different manner. The aim of this approach is to model
the interaction between the four core fields sketched in this diagram
by a {\em state space approach}. This requires to identify relevant
variables, cf.\ the diagram, by establishing functional relationships
between them in form of an equation 
\index{state space}
\be
{\bf u}_{t+1}=F({\boldsymbol\Omega},{\boldsymbol\Phi},{\bf r}_t,{\bf
    w}_t,{\bf u}_t) \label{eq:mepssa}\:,
\ee
where ${\bf u}_{t}$ is the location of an organism at time $t$. A
simple, boiled-down example of such an equation is the Langevin
equation Eq.~(\ref{eq:ule}) that we proposed to describe bumblebee
flights under predation threat. Here $du_{t+1}/dt=v_y(t)$ and the
potential term is related to the variable $r_t$ above while all the
other variables are ignored.

\section{Conclusions}

The discussion about the LFH is still very much ongoing. As an example
we refer to research on movements of mussels, where experimental
measurements seemed to suggest that L\'evy movement accelerates
pattern formation \cite{dJWH11}; however, see the discussion that
emerged about these findings as comments and replies to the above
paper, which mirrors our discussion in the previous sections. A second
example is the debate about a recent review by Andy Reynolds
\cite{Reyn15}, in which yet another new version of a LFH was
suggested; again, see all the respective comments and the authors'
reply to them. While these two articles are in support of the LFH, we
refer to a recent review by Graham Pyke \cite{Pyke15} as an example of
a more critical appreciation of it.
  
\index{power law} We conclude that one needs to be rather careful with
following power law hypotheses, or paradigms, for data analysis, here
applied to the problem of understanding the search for food by
biological organisms.  These laws are very attractive because of their
simplicity, and because in certain physical situations they represent
underlying universalities. While they clearly have their justification
in specific settings, these are rather simplistic concepts that ignore
many details of the biological situation at hand. This can cause
problems when biological processes are more complex. What we have
outlined represents not an entirely new scientific lesson; see, e.g.,
the discussion about power laws in self-organized criticality. On the
other hand, the LFH did pioneer a new way of thinking that goes beyond
applying simple traditional random walk schemes to understand
biological foraging.

Financial support of this research by the MPIPKS Dresden and the
Office of Naval Research Global is gratefully acknowledged.

\printindex
\end{document}